\documentclass[conference]{IEEEtran}
\IEEEoverridecommandlockouts
\usepackage{cite}
\usepackage{amsmath,amssymb,amsfonts}
\usepackage{algorithmic}
\usepackage{graphicx}
\usepackage{textcomp}
\usepackage{xcolor}
\def\BibTeX{{\rm B\kern-.05em{\sc i\kern-.025em b}\kern-.08em
    T\kern-.1667em\lower.7ex\hbox{E}\kern-.125emX}}
\begin{document}

\title{Channel Modeling and Multi-User Precoding for Tri-Polarized Holographic MIMO Communications  
}

\author{\IEEEauthorblockN{Li Wei$^{1}$, Chongwen Huang$^{2}$, George~C.~Alexandropoulos$^{3,4}$, Zhaohui Yang$^{2}$, \\ Jun Yang$^{5,6}$, Wei E. I. Sha$^{2}$, M\'{e}rouane~Debbah$^{4,7}$, 
		and Chau Yuen$^{1}$}
$^{1}$Singapore University of Technology and Design, Singapore\\
$^{2}$College of Information Science and Electronic Engineering, Zhejiang University, China\\
$^{3}$Department of Informatics and Telecommunications, National and Kapodistrian University of Athens, Greece\\
$^{4}$Technology Innovation Institute, Abu Dhabi, United Arab Emirates\\
$^{5}$State Key Laboratory of Mobile Network and Mobile Multimedia Technology, Shenzhen, China \\
$^{6}$Wireless Product R\&D Institute, ZTE Corporation, Shenzhen, China  \\
$^{7}$CentraleSupelec, University Paris-Saclay, Gif-sur-Yvette, France
} 

\maketitle

\begin{abstract}
This paper studies the exploitation of triple polarization (TP) for multi-user (MU) holographic multiple-input multiple-output surface (HMIMOS) wireless communication systems, aiming at capacity boosting without enlarging the antenna array size. We specifically consider that both the transmitter and receiver are equipped with an HMIMOS comprising compact sub-wavelength TP patch antennas. To characterize TP MU-HMIMOS systems, a TP near-field channel model is proposed using the dyadic Green's function, whose characteristics are leveraged to design a user-cluster-based precoding scheme for mitigating the cross-polarization and inter-user interference contributions. A theoretical correlation analysis for HMIMOS with infinitely small patch antennas is also presented. According to the proposed scheme, the users are assigned to one of the three polarizations, which is easy to implement, at the cost, however, of reducing the system's diversity. Our numerical results showcase that the cross-polarization channel components have a non-negligible impact on the system performance, which is efficiently eliminated with the proposed MU precoding scheme.  
\end{abstract}


\section{Introduction}
Holographic multiple-input multiple-output surfaces (HMIMOSs) for serving multi-user (MU) wireless communication systems are lately gaining remarkable attention as efficient transceiver front-ends for intelligently controlling electromagnetic (EM) signal propagation \cite{9136592, 9779586,9765526}. An HMIMOS consists of almost infinite antennas of compact sizes, leveraging the propagation characteristics of EM waves \cite{RISE6G_COMMAG}, and it is verified to be competitive in many fields, including unmanned aerial vehicles, millimeter waves, and multi-antenna systems \cite{yuan2022recent}. However, the full exploitation of this technology is still infeasible due to various non-trivial technical issues.   

The main challenge of the HMIMOS technology is the performance gain that depends on the array size. Since an HMIMOS incorporates large amount of patch antennas in a small area with inter-element spacing less than half of the wavelength, there exists strong correlation between the patch antennas, which degrades the performance. In fact, it has been proved that the degrees of freedom brought by increasing the number of patch antennas are limited by the size of the HMIMOS \cite{9650519}. Hence, it is still unknown how to effectively improve the spectral efficiency of an HMIMOS of a given area, when its performance limit is reached. 

The integration of the dual-polarization (DP) or tri-polarization (TP) feature is expected to further improve the performance without enlarging antenna array size, offering polarization diversity which can boost spectral efficiency \cite{9440813}. Therefore, a few recent works discussed the deployment of polarized reconfigurable intelligent surfaces (RISs) \cite{9475466,9339948}. A DP RIS-based transmission system to achieve low-cost ultra-massive MIMO transmission was designed in \cite{9475466}. In \cite{9339948}, an RIS-based wireless communication structure to control the reflected beam and polarization state for maximizing the received signal power was proposed. Nevertheless, there are still open challenges with polarized wireless communications.

The primary difficulty of wireless systems using polarization is channel modeling. Different from conventional channel models, the polarized channel involves the co-polarized and cross-polarized channels. Besides, for polarized channel, there exists an interplay between spatial and polarization correlation \cite{8299445}. Thus, the conventional independent and identically distributed assumption cannot be directly adopted to model co-/cross-polarized channels \cite{9497725}. In addition, polarized wireless communications suffer from power imbalance \cite{9413660}, and the power allocation becomes the major concern \cite{xu2022resource}.  However, in most of the existing literature, the power of co-polarized channels is assumed to be equal. Therefore, a realistic channel model including polarization is required. 

In order to exploit the potential of polarized HMIMOS communications, this paper presents a TP MU-HMIMOS channel model for the near-field (NF) regime, which is deployed for designing a user-cluster-based precoding scheme aiming to mitigate inter-user and cross-polarization interference. The proposed channel model is proved to be efficient enough through our theoretical analysis, and via numerical results, it is showcased that TP HMIMOS systems are more efficient than DP and conventional HMIMOS systems. 

The remainder of this paper is organized as follows. In Section~\ref{sec:signal model}, the TP MU-HMIMOS system model is introduced. The near-field channel model for TP MU-HMIMOS systems and its theoretical analysis are presented in Section~\ref{sec:channel_model}. Section~\ref{sec: UE_precode} presents a user-cluster-based precoding scheme, while Section~\ref{sec:simulation} shows our numerical investigations. Finally, the paper's conclusions are drawn in Section~\ref{sec:conclusion}. 

\section{System Modeling}\label{sec:signal model}
\begin{figure} [!t]
	\begin{center}
		\centerline{\includegraphics[width=0.5\textwidth]{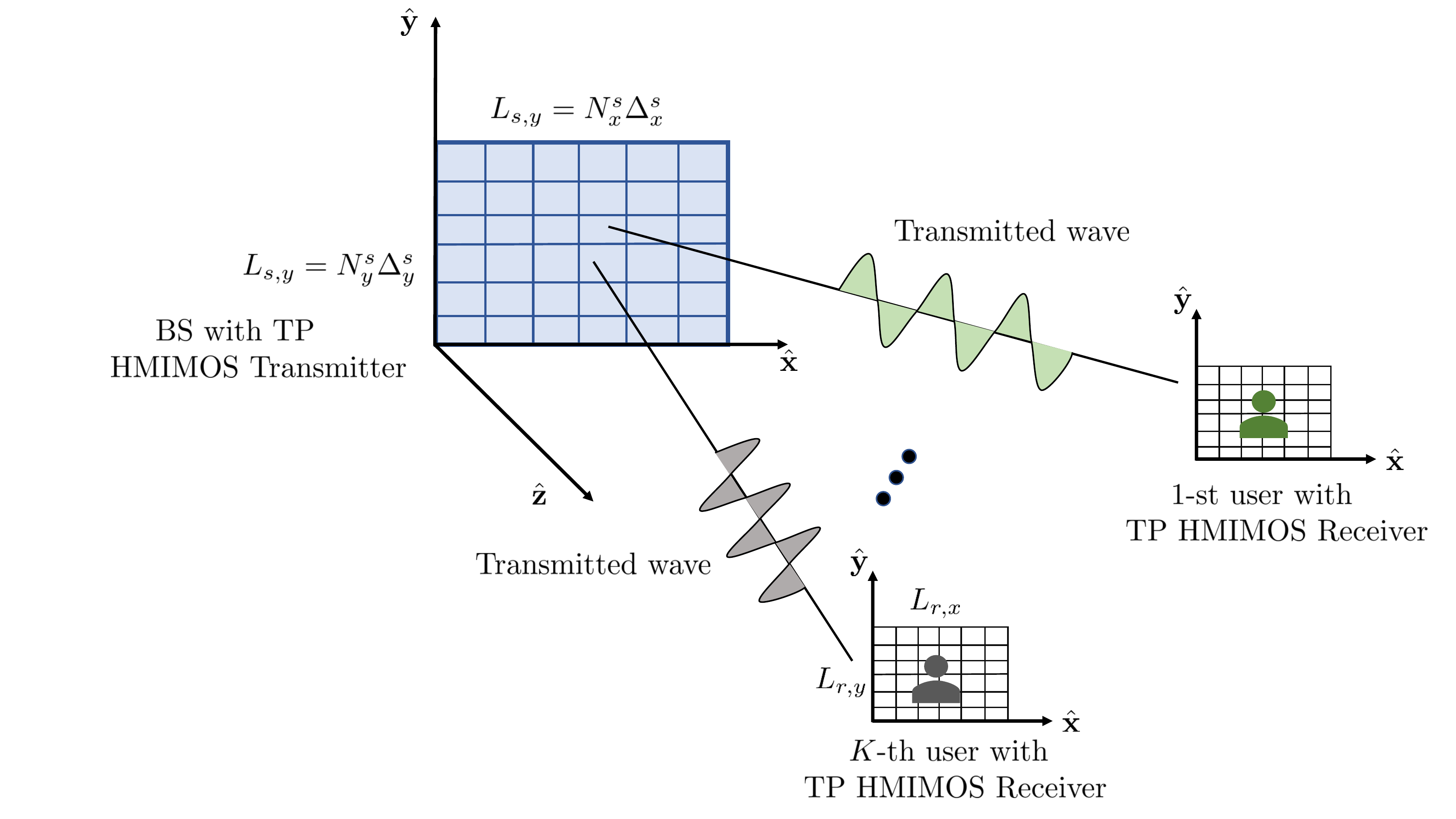}}  
		\caption{The considered TP MU-HMIMOS downlink communication system consisting of a BS with $N_s$ patch antennas serving $K$ users each equipped with $\bar{N}_r$ patch antennas.}
		\label{fig:SystemModel} 
	\end{center}
\end{figure}

In the considered TP MU-HMIMOS communication system, the BS is equipped with an HMIMOS of size $A_s=L_{s,x} \times L_{s,y}$  where $L_{s,x}$ and $L_{s,y}$ denote surfaces'  horizontal and vertical lengths, respectively, and the $K$ users have also a TP HMIMOS of size $A_r=L_{r,x} \times L_{r,y}$ with $L_{r,x}$ and $L_{r,y}$ being their common horizontal and vertical lengths, respectively. The HMIMOSs at the BS and each of the users are composed of $N_s$ and $\bar{N}_r$ patch antennas, respectively. Thus, the sum of the receive patch antennas in the downlink is $N_r=K \bar{N}_r$. Each patch antenna is made from metamaterials that are capable of independently adjusting their reflection coefficients in three polarizations \cite{9475466,9765815,9324910}, as shown in Fig$.$~\ref{fig:SystemModel}. We represent the phase configuration matrix (i.e., the analog beamforming) of the HMIMOS at the BS by the block-diagonal matrix $\mathbf{\Phi}^s\in\mathbb{C}^{3N_s\times 3N_s}$, whose non-zero blocks $\mathbf{\Phi}_n^s\in \mathbb{C}^{3\times 3}$, with $n=1,2,\ldots,N_s$, refer to the three-polarization configuration of each $n$-th patch antenna. Each of these configurations is represented by the following diagonal matrix:
\begin{equation}
	\mathbf{\Phi}_{n}^s=\mathrm{diag}[
		E_{n,x}^s e^{j \theta_{n, x}^s},E_{n,y}^s e^{j \theta_{n, y}^s},E_{n,z}^s e^{j \theta_{n, z}^s}],
\end{equation}
where $\theta_{n, x}^s, \theta_{n, y}^s$, and $\theta_{n, z}^s\in [0,2\pi]$ are three independent phase shifts in the excited triple polarization states, with \ $E_{n,x}^s, E_{n,y}^s$, and $E_{n,z}^s \in [0,1]$ denoting the corresponding amplitude reflection coefficients at each $n$-th patch antenna.  

Assuming that the EM wave propagates towards the $\hat{\mathbf{z}}$ direction, thus, it can be decomposed into the components $(E_x,E_y,E_z)$ in the three orthogonal directions, i.e., in $\hat{\mathbf{x}}$, $\hat{\mathbf{y}}$, and $\hat{\mathbf{z}}$ directions, the instantaneous electrical field is given by 
\vspace{-0.1cm}
\begin{equation}
	\mathbf{E}(t)=E_x(t) \hat{\mathbf{x}} +E_y (t) \hat{\mathbf{y}} +E_z(t) \hat{\mathbf{z}},
\end{equation}
\vspace{-0.1cm}
where  $E_{j}(t)\triangleq E_{j}  \exp(i(\omega t+\theta_{n,i}))$ with $j\in\{x,y,z\}$, and $(\theta_{n,x} ,\theta_{n,y} ,\theta_{n,z} )$ are the phases of the three polarized components that are controlled by each $n$-th patch antenna in the HMIMOS. In the sequel, the term referring to the dependence in time is omitted for simplicity.

\section{Proposed Near-Field Channel Model} \label{sec:channel_model}
The radiated electric field $\mathbf{E}(\mathbf{r})$ at the location $\mathbf{r}\in\mathbb{R}^{3}$ in the half free-space, resulting from the current $\mathbf{J}(\mathbf{r}')$, which is generated at the location $\mathbf{r}'\in\mathbb{R}^{3}$, is given by the dyadic Green's function theorem as follows \cite{5991926}:
\vspace{-0.1cm}
\begin{equation}\label{eq:elec_field}
	\mathbf{E}(\mathbf{r})\triangleq  i \omega \mu \int_{S} d s^{\prime}  {\bar{\mathbf{G}}}\left(\mathbf{r}, \mathbf{r}^{\prime}\right) \mathbf{J}\left(\mathbf{r}^{\prime}\right),
\end{equation}
\vspace{-0.1cm}
where $S$ denotes the surface of the HMIMOS transmitter, $\omega$ is permittivity, and $\mu$ is permeability. The dyadic Green's function is defined as \cite{arnoldus2001representation}:
\vspace{-0.1cm}
\begin{equation} \label{equ:Green's function}
	\begin{aligned}
	&{\bar{\mathbf{G}}}\left(\mathbf{r}, \mathbf{r}^{\prime}\right)\triangleq\left[\bar{\mathbf{I}}+\frac{\nabla \nabla}{k_0^{2}}\right] g\left(\mathbf{r}, \mathbf{r}^{\prime}\right)\\
	 & = \!\!\left(\!1\!\!+\!\!\frac{i}{k_0 r }\!\!-\!\!\frac{1}{k_0^{2} r ^{2}}\!\right) \!\bar{\mathbf{I}} g(\mathbf{r} \!,\!\mathbf{r}' )\! \!+\!\!\left(\!\frac{3}{k_0^{2} r ^{2}}\!\!-\!\!\frac{3 i}{k_0 r }\!\!-\!\!1\!\right) \vec{\mathbf{r}}  \vec{\mathbf{r} }    g(\mathbf{r}\! ,\!\mathbf{r}' ),
	\end{aligned}
\end{equation}
\vspace{-0.1cm}
where $\bar{\mathbf{I}}$ is identity matrix, $\nabla \nabla g (\cdot)$ denotes the second-order derivative of function $g(\cdot)$ with respect to its argument, $k_0\triangleq\frac{2\pi}{\lambda}$ is the wavenumber with $\lambda$ being the wavelength, and the unit vector $\vec{\mathbf{r}}$ denotes the direction between the source point and radiated field. The scalar Green's function is  \cite{arnoldus2001representation}:
\vspace{-0.1cm}
\begin{equation}
	g\left(\mathbf{r}, \mathbf{r}^{\prime}\right)\triangleq\frac{e^{i k_0\left|\mathbf{r}-\mathbf{r}^{\prime}\right|}}{4 \pi\left|\mathbf{r}-\mathbf{r}^{\prime}\right|}.
\end{equation}
\vspace{-0.1cm}

Inspired by the work in \cite{19157}, which calculates the sound field radiated from a plane source using the scalar Green's function, we derive the dyadic Green's function that takes polarization into consideration. Specifically, the $N_s$-patch HMIMOS is divided into $N_s$ rectangles, each with size $ \Delta^s = \Delta_x^s \Delta_y^s$. Each patch is regarded as a point $(x'_n,y'_n)$ in a first coordinate system, and is further investigated in a second coordinate system defined within $(x'_0,y'_0)$ (this is equivalent to the size of each patch antenna). Under this consideration, the electric field  can be rewritten as follows:
\vspace{-0.1cm}
\begin{equation}\label{eq:field}
	\begin{aligned}
		&\mathbf{E}(\mathbf{r})=  \sum_{n=1}^{N_s} \int \int \mathrm{d} x^{\prime}_n\mathrm{d} y^{\prime}_n  {\bar{\mathbf{G}}}\left(\mathbf{r}, \mathbf{r}^{\prime}_n\right) \mathbf{J}\left(\mathbf{r}^{\prime}_n\right)\\ 
		&=\sum_{n=1}^{N_s} \!\int_{-\Delta_x^s/2}^{\Delta_x^s/2} \!\int_{-\Delta_y^s/2}^{\Delta_y^s/2} d x'_0 d y'_0  \left[\bar{\mathbf{I}}\!+\!\frac{\nabla \nabla}{k_0^{2}}\right] \frac{e^{i k_0 r_n }}{4 \pi r_n }  \! \mathbf{J}\left(\mathbf{r}^{\prime}_n\right),
	\end{aligned}
\end{equation}
\vspace{-0.1cm}
with the distance $r_n$ given by:
\vspace{-0.1cm}
\begin{equation}
	\begin{aligned}
		r_n=|\mathbf{r}-\mathbf{r}'_n|&=\sqrt{(\hat{x}_n-x'_0)^2 \!+\!( \hat{y}_n-y'_0)^2\! +\! z^2},
	\end{aligned}
\end{equation}
\vspace{-0.1cm}
where $\hat{x}_n\triangleq x-x'_n$ and $\hat{y}\triangleq y -y'_n$. We henceforth assume, for simplicity, that the current distribution $\mathbf{J}(\mathbf{r}'_n)=J_x(\mathbf{r}'_n) \hat{\mathbf{x}}+J_y(\mathbf{r}'_n) \hat{\mathbf{y}}$ is constant, and  $J_x(\mathbf{r}'_n) =J_y(\mathbf{r}'_n) =1$.  

If $\Delta_x^s$ and $\Delta_y^s$ are infinitely small, the distance to the field point (i.e., the receiver) is much greater than the dimensions of the source, hence, the Fraunhofer approximation can be adopted, i.e., $\sqrt{{x'_0}^2+{y'_0}^2}/{\tilde{R}_n}\approx (0,0)$, where $\tilde{R}_n$ is:
\vspace{-0.1cm}
\begin{equation}
	\tilde{R}_n\!=\!\sqrt{(x\!-\!x'_n)^2 \!+\! (y\!-\!y'_n)^2 \!+\! z^2}\!=\!\sqrt{\hat{x}_n^2 \!+\! \hat{y}_n^2\! +\! z^2}.
\end{equation}
\vspace{-0.1cm} 
In this case, the distance ${r}_n$ from each $n$-th transmitting patch antenna in the exponential term in expression \eqref{eq:field} can be approximated as follows \cite{19157}:
\vspace{-0.1cm}
\begin{equation}  \label{equ:assump1}
	\begin{aligned}
		 { \tilde{r}_n}&=  \sqrt{(\hat{x}_n-x_0)^2 + (\hat{y}_n-y_0)^2 + z^2} \\ 
		&\overset{(a)} {\approx} \!\!  \tilde{R}_n\!- \!\frac{\hat{x}_n x'_0}{\tilde{R}_n}\!  +\!\frac{{x'_0}^2}{2\tilde{R}_n } \!\- \! \frac{\hat{y}_n y'_0}{\tilde{R}_n}  \!+\! \frac{{y'_0}^2}{2 \tilde{R}_n}\! + \!\tilde{R}_n \mathcal{O} \left(u(\mathbf{r},\mathbf{r}'_n)\right)   \\
		&\overset{(b)} {\approx} \!\! \tilde{R}_n- \frac{\hat{x}_n x'_0}{\tilde{R}_n}    -  \frac{\hat{y}_n y'_0}{\tilde{R}_n}     ,
	\end{aligned}
\end{equation}
\vspace{-0.1cm}
where $(a)$ is Taylor series expansion and $u(\mathbf{r},\mathbf{r}'_n)\triangleq\frac{2\hat{x}_n x'_0}{\tilde{R}_n^2}  -\frac{ {x'_0}^{2}}{\tilde{R}_n^2 } +  \frac{2 \hat{y}_n y'_0}{\tilde{R}_n^2}  - \frac{  {x'_0}^2}{ \tilde{R}_n^2}$, with $\mathcal{O} \left(u(\mathbf{r},\mathbf{r}'_n)\right)$ being the negligible higher order terms. In addition, $(b)$ results from the fact that the term $\left|\frac{\sqrt{{x'_0}^2+{y'_0}^2} } {2\tilde{R}_n}\right|$ is small. 

By making the reasonable assumption that $1/r_n \approx 1/\tilde{R}_n$, the radiated electric field from each $n$-th transmitting patch antenna can be obtained as follows:
\vspace{-0.1cm}
\begin{equation} \label{equ:assump2}
	\begin{aligned}
		&\mathcal{E}(\tilde{R}_n) =\sum_{n=1}^{N_s} \int_{-\Delta_x^s/2}^{\Delta_x^s/2} \int_{-\Delta_y/2}^{\Delta_y^s/2} d x'_0 d y'_0    \frac{e^{i k_0\left|\mathbf{r}-\mathbf{r}^{\prime}_n\right|}}{4 \pi \tilde{R}_n}  \\ 
		&= \sum_{n=1}^{N_s} \int_{-\Delta_x^s/2}^{\Delta_x^s/2} \int_{-\Delta_y^s/2}^{\Delta_y^s/2} d x'_0 d y'_0  \tilde{\mathbf{G}}  (\hat{x}_n,\hat{y}_n;x'_0,y'_0) ,
	\end{aligned}
\end{equation}
\vspace{-0.1cm}
where we have used the function definition:
\vspace{-0.1cm}
\begin{equation}
	\begin{aligned}
		& \tilde{\mathbf{G}} (\hat{x}_n,\hat{y}_n;x'_0,y'_0)  =  \frac{\exp \{i k_0 \left[\tilde{R}_n- \frac{\hat{x}_n x'_0}{\tilde{R}_n}    -  \frac{\hat{y}_n y'_0}{\tilde{R}_n}     \right]\} }{4 \pi\tilde{R}_n}\\
		&=\frac{e^{ik_0 \tilde{R}_n}}{4 \pi \tilde{R}_n}   {\exp\left(ik_0 \left[- \frac{\hat{x}_n x'_0}{\tilde{R}_n}       \right]\right)}    {\exp \left(i k_0 \left[  -  \frac{\hat{y}_n y'_0}{\tilde{R}_n}     \right] \right) } .
	\end{aligned}
\end{equation}
\vspace{-0.1cm}
Using the notation $\Delta^s=\Delta_x^s \Delta_y^s$ and $\mathrm{sinc}(x)=\frac{\mathrm{sin} x}{x}$, the last integral in \eqref{equ:assump2} can be solved as follows:
\vspace{-0.1cm}
\begin{equation}
	\begin{aligned}
		&\mathcal{E}(\tilde{R}_n) \!  = \!\sum_{n=1}^{N_s}\! \int_{-\Delta_x/2}^{\Delta_x/2} \!\int_{-\Delta_y^s/2}^{\Delta_y^s/2} \!d x'_0 d y'_0  \tilde{\mathbf{G}}  (\hat{x}_n,\hat{y}_n;x'_0,y'_0)  \\
		&= \Delta^s \sum_{n=1}^{N_s} \frac{e^{ik_0 \tilde{R}_n}}{4 \pi \tilde{R}_n}    \operatorname{sinc} \frac{k_0 \hat{x}_n  \Delta_x^s}{2 \tilde{R}_n}  \operatorname{sinc} \frac{k_0 \hat{y}_n  \Delta_y^s}{2 \tilde{R}_n} .
	\end{aligned}
\end{equation}  
\vspace{-0.1cm}
By making use of the following two terms given in \eqref{equ:Green's function}:
\begin{equation} \label{equ:GreenVector_term}
	\begin{aligned}
	\!c_1(\!\tilde{R}_n\!)\!\!=\!\!\left( \!1\!\!\!+\!\!\frac{i}{k_0 \tilde{R}_n}\!\!-\!\!\frac{1}{k_0^{2}\! \tilde{R}_n^{2}}\!\right)\!, c_2(\!\tilde{R}_n\!)\!\!=\!\!\left(\!\frac{3}{k_0^{2} \tilde{R}_n^{2}}\!\!-\!\!\frac{3 i}{k_0 \!\tilde{R}_n}\!\!-\!\!1\!\!\right)\!,
	\end{aligned}
\end{equation}
the wireless channel with polarization between each $n$-th transmit patch antenna and a receiving point can be represented by
\vspace{-0.1cm}
\begin{equation}
	\begin{aligned}
		\mathbf{H}_{n}= \mathcal{E}(\tilde{R}_n) \mathbf{C}_n = \mathcal{E}(\tilde{R}_n)	\left(c_1(\tilde{R}_{n}) \mathbf{I}+c_2(\tilde{R}_{n})  \vec{\mathbf{r}}_{n} \vec{\mathbf{r} }_{n}^{T}\right)  , 
	\end{aligned}
\end{equation}
\vspace{-0.1cm}
where the unit vector $\vec{\mathbf{r}}_n\triangleq\frac{\mathbf{r}-\mathbf{r}'_n}{r_n}$ denotes the direction of each receive-transmit patch-antenna pair.

\subsection{Channel Matrix and Feasibility}
It is, in general, expected that the receive HMIMOS will be much smaller than the transmit one, hence, it is reasonable to assume that the power received by each patch antenna will be proportional to the receive area $\Delta^r\triangleq\Delta^r_x \Delta^r_y$. Therefore, the channel between each $m$-th, with $m=1,2,\ldots,\bar{N}$, receive and each $n$-th transmit patch antennas can be expressed as:
\vspace{-0.1cm}
\begin{equation} \label{equ:CM_fullPolar}
	\begin{aligned}
		\mathbf{H}_{mn} =& \Delta^s  \Delta^r \frac{\exp\left(ik_0 \tilde{R}_{mn}\right)}{4 \pi \tilde{R}_{mn}}    \operatorname{sinc} \frac{k_0 (x_m-x_n')  \Delta^s_x}{2 \tilde{R}_{mn}}\\
		&\times  \operatorname{sinc} \frac{k_0 (y_m-y_n')  \Delta^s_y}{2 \tilde{R}_{mn}} \mathbf{C}_{mn}   \in \mathbb{C}^{3 \times 3},
	\end{aligned}
\end{equation} 
\vspace{-0.1cm}
where $\mathbf{C}_{mn}\triangleq c_1(\tilde{R}_{mn}) \mathbf{I}+c_2(\tilde{R}_{mn})  \vec{\mathbf{r}}_{mn} \vec{\mathbf{r} }_{mn}^{T} \in \mathbb{C}^{3\times 3}$. The overall channel matrix can be thus represented as follows:
\vspace{-0.1cm}
\begin{equation}  \label{equ:polarized_H}
	\begin{aligned}
		\mathbf{H} & = \left[\begin{array}{ccc }
			\mathbf{H}_{xx}  & \mathbf{H}_{xy}& \mathbf{H}_{xz} \\ 
			\mathbf{H}_{yx}  & \mathbf{H}_{yy}& \mathbf{H}_{yz} \\
			\mathbf{H}_{zx}  & \mathbf{H}_{zy}& \mathbf{H}_{zz} 
		\end{array}\right]\in \mathbb{C}^{3N_r \times 3 N_s},
	\end{aligned}
\end{equation}
\vspace{-0.1cm}
where $\mathbf{H}_{pq}\in\mathbb{C}^{ N_r \times  N_s}$, with $p,q\in\{x,y,z\}$, denotes the polarized channel that collects all channel components transmitted in the $p$-th polarization and received in the $q$-th polarization. We will next prove the validity of our two core assumptions following our proposed channel model. 

Recall that, in order to derive the above expressions, we have used the following two approximations:
\begin{itemize}
	\item  $\exp \left(i k \left[  -  \frac{\hat{x}_n \Delta_x}{2\tilde{R}_n} -  \frac{\hat{y}_n \Delta_y}{2\tilde{R}_n} \right] \right) \approx 1$ (or equivalently $\left|\frac{\mathbf{r}'_0}{ 2 \tilde{R}_n} \right| \approx 0$ ) was adopted in \eqref{equ:assump1}, implying that the transmitter patch antenna is small enough compared to twice the distance between the transmitter and receiver;  
	\item $ r_n \approx  \tilde{R}_n$ was used in the derivation of \eqref{equ:assump2}.
\end{itemize}
Following these assumptions, the length limit of each patch antenna can be deduced. Specifically, since $\exp\left( \frac{ik_0 ({x'_0}^2 +{y'_0}^2)}{2\tilde{R}_{n}}\right)\!\approx\! 1$, $-\Delta_x^s\!\leq \! x'_0\leq\! \Delta_x^s$, and $-\Delta_y^s\!\leq \! y'_0\!\leq \!\Delta_y^s$, it holds $\cos \left( \frac{ k_0 ({\Delta_x^s}^2 +{\Delta_y^s}^2)}{8\tilde{R}_{n}}\right)\approx 1$. Thus, $\frac{ k_0 ({\Delta_x^s}^2 +{\Delta_y^s}^2)}{8\tilde{R}_{n}}\ll \pi$, or equivalently, $ \frac{k_0 {\Delta_x^s}^{2}}{ 8 \tilde{R}_n} \ll \pi$ and $\frac{k_0 {\Delta_y^s}^{2}}{ 8 \tilde{R}_n}  \ll \pi$. Therefore, the limitation on the patch antenna sizes at the receiver and transmitter can be explicitly given by 
\vspace{-0.1cm}
\begin{equation}
	\Delta_x^r\leq\Delta_x^s\ll 2 \sqrt{\lambda \tilde{R}_n}, \quad \Delta_y^r\leq\Delta_y^s\ll 2 \sqrt{\lambda \tilde{R}_n}m.
\end{equation}
\vspace{-0.1cm}

Since the near-field region is determined by  condition $r_{\mathrm{NF}}\leq\frac{2(D_1+D_2)^2}{\lambda}$ \cite{1451288, cui2022near}, the aperture at the transmitter is $D_1=\sqrt{L_{s,x}^2+L_{s,y}^2}$ and the aperture at the receiver is $D_2=\sqrt{L_{r,x}^2+L_{r,y}^2}$. If we consider that $N_{s,x}=N_{s,y}=\sqrt{N_s}$ ($N_{s,x}$ and $N_{s,y}$ are the numbers of horizontal and vertical patch antennas at the BS) and $N_{r,x}=N_{r,y}=\sqrt{N_r}$ ($N_{r,x}$ and $N_{r,y}$ denote the numbers of the horizontal and vertical patch antennas at each user), the near-field region is deduced:
\vspace{-0.1cm}
\begin{equation}
	r_{\mathrm{NF}}\leq\frac{ 4N_s(\Delta^s_x)^2+4N_r(\Delta^r_x)^2 +8  \sqrt{N_sN_r} \Delta^s_x\Delta^r_x }{\lambda}.
\end{equation}
\vspace{-0.1cm}
By letting $\tilde{R}_n=r_{\mathrm{NF}}$, the length limit of each patch antenna becomes as follows:
\vspace{-0.1cm}
\begin{equation}
	\begin{aligned}
		&(\Delta_x^s)^2+(\Delta_y^s)^2 \ll   4 {\lambda \tilde{R}_n}\\
		&=16 ( N_s(\Delta^s_x)^2+ N_r(\Delta^r_x)^2 +2  \sqrt{N_sN_r} \Delta^s_x\Delta^r_x ).  
	\end{aligned}		
\end{equation}
\vspace{-0.1cm}
Note that this inequality always holds since $16 N_s\gg 1$ and $16 N_r \gg 1$, which proves the feasibility of the proposed channel model for TP HMIMOS communication systems.  

\subsection{Correlation Analysis} 
There are mainly two channel correlation matrices: the correlation matrix $\mathbf{R}^s$ at the transmitter and the correlation matrix $\mathbf{R}^r$ at each receiver. The $n$-th transmit patch antenna with location $\mathbf{r}_{n}$, and the $n'$-th transmit patch antenna with location $\mathbf{r}_{n'}$, will be spatially correlated by the following factor:
\vspace{-0.1cm}
\begin{equation}\label{eq:corr}
	\begin{aligned}
		&\mathbf{R}^s\left(\mathbf{r}_{n}, \mathbf{r}_{n'} \right) \propto	\left\langle \mathbf{E} \left(\mathbf{r}_{n}\right) \mathbf{E}^{\dagger}\left(\mathbf{r}_{n'}\right)\right\rangle  \\
		&= \! \int\!\!\!\!\int \!\!  d^2 \mathbf{r}_{m} d^2 \mathbf{r}_{m'}  {\bar{\mathbf{G}}}\!\left(\!\mathbf{r}_m, \!\mathbf{r}_n\right) \!{\bar{\mathbf{G}}^{\dagger}}\!\!\left(\!\mathbf{r}_{m'}, \!\mathbf{r}_{n'}\right) \!\!\left\langle \mathbf{J}(\mathbf{r}_m) \mathbf{J}(\mathbf{r}_{m'} ) \! \right\rangle\!, 
	\end{aligned}
\end{equation}
\vspace{-0.1cm}
where $\dagger$ denotes transpose operation.
Due to the reciprocity, the receive correlation can also be obtained in the same way.  We assume that $\left\langle \mathbf{J}(\mathbf{r}_m) \mathbf{J}(\mathbf{r}_{m'} ) \right\rangle=\delta_{mm'}/3$, as a consequence of the fluctuation-dissipation theorem \cite{HENKEL200057,Chew_PIER}, we have:
\vspace{-0.1cm}
\begin{equation} 
	\begin{aligned}
		&\mathbf{R}^s \left(\mathbf{r}_{n},\mathbf{r}_{n'}   \right) \propto \mathrm{Im}\{\bar{\mathbf{G}}(d_{nn'})\}, 
	\end{aligned}
\end{equation}
\vspace{-0.1cm}
where $d_{nn'}$ is the distance between the $n$th and $n'$th transmit patch antennas. 
 
Thus, the imaginary part of Green's in the $x$th polarization is given by (the constant term $\frac{1}{4\pi}$ is omitted here):
\vspace{-0.1cm}
\begin{equation}
	\begin{aligned}
		&\mathrm{Im} \{{\bar{\mathbf{G}}}(d_{nn'}\!)_{x}\}  \!\!=\!\!\frac{\sin (k_0 d_{n\!n'})}{  d_{n\!n'}} \!\!+\! \!\frac{\cos (k_0 d_{n\!n'}\!)}{  k_0 d_{n\!n'}^2}\!\! -\!\! \frac{\sin (k_0 d_{n\!n'}\!)}{  k_0^2 d_{n\!n'}^3}  \\
		&\quad+ \! \!\frac{3 \check{x}^2 \!\sin(\!k_0 d_{nn'}\!)}{k_0^{2} d_{nn'} ^{5}}-\!\!\frac{3\check{x}^2  \!\cos (\!k_0 d_{nn'}\!)}{k_0 d_{nn'}^4 }\!\!- \!\!\frac{\check{x}^2 \! \sin (\!k_0 d_{nn'}\!)}{  d_{nn'}^3}\\
		&\overset{(a)}{=}  {k_0} +\frac{1 }{  k_0 r^2} -\frac{  k_0   }{2 }- \frac{1 }{  k_0  r^2} +\frac{  k_0  }{6 }  	+\frac{3\check{x}^2     }{ k_0  r^4  } - \frac{ \check{x}^2   k_0   }{ 2  r^2   }\\
		&\quad	-\frac{3\check{x}^2       }{ k_0 r^4  }+\frac{3\check{x}^2   {  k_0  }   }{2  r^2  }-\frac{ \check{x}^2   { k_0^3   } }{8  }  - \frac{\check{x}^2 k_0   }{  r^2 } +\frac{\check{x}^2   { k_0^3  }  }{6   }\\
		& =\frac{ 2 k_0  }{3 }    +\frac{\check{x}^2   { k_0^3   }  }{24 },
	\end{aligned}
\end{equation}
\vspace{-0.1cm}
where $\check{x}=x-x'$ being the $x$-coordinate difference between the  $n$th and $n'$th transmit patch antennas. (a) is obtained by Taylor's series expansion, which is feasible in HMIMOS with infinitely small patch antennas. 
%
%

Therefore, 
\vspace{-0.1cm}
\begin{equation} \label{equ:CorrelationTaylor}
	\begin{aligned}
		&\frac{\mathrm{Im} \{{\bar{\mathbf{G}}}( d_{nn'})_{x}\} + \mathrm{Im} \{{\bar{\mathbf{G}}}( d_{nn'})_{y}\}  + \mathrm{Im} \{{\bar{\mathbf{G}}}( d_{nn'})\}_{z}}{12\pi}\\
		&=\frac{  k_0  }{6\pi} +\frac{ { k_0^3  }  }{24  } \cdot \frac{\check{x}^2  +\check{y}^2  +\check{z}^2  }{12\pi}=\frac{   k_0  }{6\pi } +\frac{ { k_0^3  d_{nn'}^2 }  }{288\pi }.  
	\end{aligned}
\end{equation}
\vspace{-0.1cm}

If $n=n'$, we have the reference field correlation
\vspace{-0.1cm}
\begin{equation} 
	\begin{aligned}
		&\mathbf{R}^s_{n }\left(\mathbf{r}_{n},\mathbf{r}_{n'}  \right) \triangleq	\left\langle \mathbf{E} \left(\mathbf{r}_{n}\right)\mathbf{E} ^{\dagger}\left(\mathbf{r}_{n'}\right)\right\rangle =\frac{   k_0  }{6\pi } .
	\end{aligned}
\end{equation}
\vspace{-0.1cm}

The above equation provides a theoretical analysis for HMIMOS with infinitely small patch antennas in an ideal scenario. It is shown that the smaller distance $d_{nn'}$ experiences the smaller spatial correlation, while the patch antenna located at larger distance undergoes larger spatial correlation. This is reasonable since the EM fields are quasi-static in the near-field, and spatial coherence occurs in overlapped source areas of the order of $\pi z^2$ \cite{HENKEL200057}.  

A similar derivation can be applied to the receiver correlation matrix $\mathbf{R}^r$. It can be concluded that, the smaller the spacing and the larger distance are between the transmitter and receiver, the larger becomes the spatial correlation. It should be noted that the user correlation is not considered in this subsection, however, a similar analysis is straightforward. 

\section{User-Cluster-Based Precoding} \label{sec: UE_precode}
There are mainly two interference sources in TP MU-HMIMOS systems: one from the cross-polarization and the other from the inter-user interference. To eliminate two interference sources, a precoding scheme based on user clustering is presented in this paper.

Based on the proposed channel model $\mathbf{H}$ in \eqref{equ:polarized_H}, the input-output relationship of the considered TP HMIMOS communication system is given by
\vspace{-0.1cm}
\begin{equation}
	\begin{aligned}
		\mathbf{y} =\!\!\left[\begin{array}{ccc }
			{\mathbf{H}}  _{xx} &  {\mathbf{H}} _{xy} &  {\mathbf{H}} _{xz} \\ 
			{\mathbf{H}} _{yx}  &  {\mathbf{H}} _{yy} &  {\mathbf{H}} _{yz} \\
			{\mathbf{H}} _{zx} &  {\mathbf{H}} _{zy} &  {\mathbf{H}} _{zz}
		\end{array}\right] \left[\begin{array}{c  }
			{\mathbf{x}}_{x}    \\ 
			{\mathbf{x}}_{y} \\
			{\mathbf{x}}_{z} 
		\end{array}\right]+\mathbf{n},
	\end{aligned}
\end{equation}
\vspace{-0.1cm}
where $\mathbf{x}\triangleq[\mathbf{x}_{x}^{T},\mathbf{x}_{y}^{T},\mathbf{x}_{z}^{T} ]^{T}\in\mathbb{C}^{3 N_r \times 1}$, with $\mathbf{x}_{p}\in\mathbb{C}^{  N_r \times 1}$ for $p\in\{x,y,z\}$, is the transmitted signal, $\mathbf{y}\in\mathbb{C}^{3 N_r \times 1}$ is the received signal, and $\mathbf{n}\in\mathbb{C}^{3 N_r \times 1}$ denotes the additive white Gaussian noise. It can be observed that the received signal suffers from the interference caused by the two cross-polarization components, which is more severe than the interference experienced in DP systems. Therefore, in order to suppress the cross-polarization interference, the precoding matrices $\mathbf{P}_x,\mathbf{P}_y$, and $\mathbf{P}_z$ are introduced, as follows:
\vspace{-0.1cm}
\begin{equation}
	\begin{aligned}
		&\!\!\mathbf{y}^{(1)}\!\! = \! \!{\mathbf{H}} _{xx}\! \mathbf{P}\!_x \!\mathbf{x}_{x}\!\! +\! \!{\mathbf{H}} _{x\! y} \!\mathbf{P}\!_y \!\mathbf{x}_{y} \!\!+\!\! {\mathbf{H}} _{x\! z} \!\mathbf{P}\!_z \!\mathbf{x}_{z}, 
		\mathbf{y}^{(2)}\!\!=\! \! {\mathbf{H}} _{y\! x} \!\mathbf{P}\!_x \!\mathbf{x}_{x} \!
	\!+\!\! {\mathbf{H}} _{y\! y} \!\\	& \!\mathbf{P}\!_y\! \mathbf{x}_{y} \!\!+\! \!{\mathbf{H}} _{y\! z}\!\mathbf{P}\!_z \!\mathbf{x}_{z},  
		\mathbf{y}^{(3)}\!\!=\!\!{\mathbf{H}} _{z\! x}\! \mathbf{P}\!_x \mathbf{x}_{x}\! \!+\!\!{\mathbf{H}} _{z\! y}\! \mathbf{P}\!_y \mathbf{x}_{y}\!\! +\!\!{\mathbf{H}} _{z\! z}\!\mathbf{P}\!_z \mathbf{x}_{z}.
	\end{aligned}
\end{equation} 
\vspace{-0.1cm}

\begin{figure} [!t]
	\begin{center}
		\includegraphics[width=0.4\textwidth]{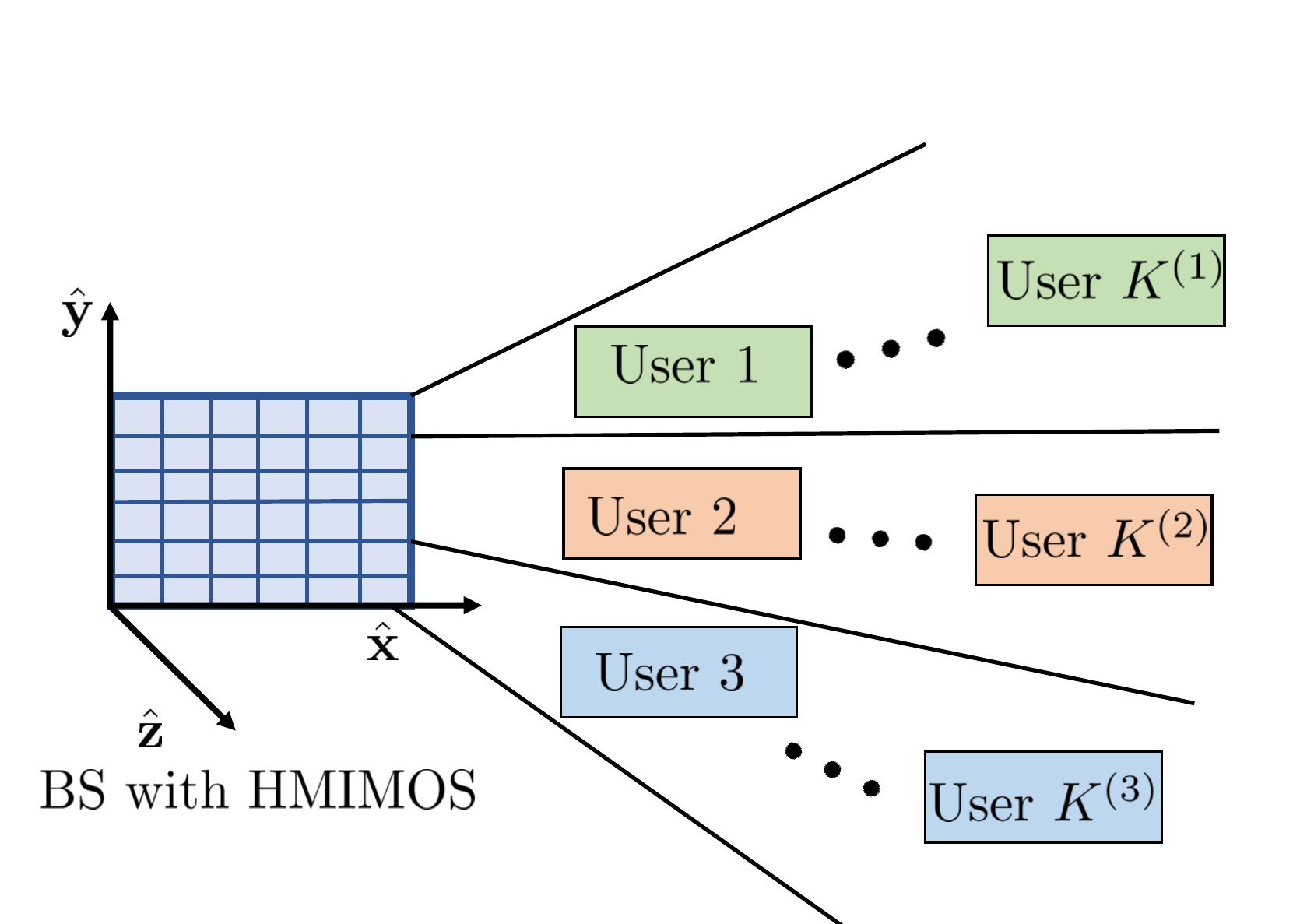}  
		\caption{The visualization of the proposed user clustering scheme for TP MU-HMIMOS systems. Each differently colored user cluster refers to a different polarization.}
		\label{fig:DP_MultiUser} 
	\end{center}
\end{figure}

Intuitively, the cross-polarization  interference can be mitigated if each user is assigned to only one polarization. Inspired by this idea, a precoding design based on user clustering was proposed in \cite{5928360} for DP communication systems. According to this scheme, each data stream can be independently transmitted in different polarizations without interference. This scheme can be also extended to TP systems, as shown in Fig.~\ref{fig:DP_MultiUser}. Specifically, the $K$ users are sorted in different polarizations based on their distances to the BS, such that $d_{1}<d_{2}<\ldots<d_{K}$, resulting in three disjoint subsets: \textit{i}) the $x$-subset $\mathcal{L}^x=\{1,4,\ldots,K^{(1)}\}$ containing $|\mathcal{L}^x|=K/3$ (we consider the case that $K/3$ is an integer) users with $K^{(1)}=K/3-2$; \textit{ii}) the $y$-subset  $\mathcal{L}^y=\{2,5,\ldots,K^{(2)}\}$ with $|\mathcal{L}^y|=K/3$ users where $K^{(2)}=K/3-1$; and \textit{iii}) the $z$-subset $\mathcal{L}^z=\{3,6,\ldots,K^{(3)}\}$ including $|\mathcal{L}^z|=K/3$ users with $K^{(3)}=K/3$. The corresponding sub-channel matrices for the user cluster are derived as:
\vspace{-0.1cm}
\begin{equation}
	\begin{aligned}
		\tilde{\mathbf{H}}_x^{\mathrm{UC}} & \!\!\!=\!\! \left[\!\!
		\begin{array}{c}
			{h}_{1 1}^{(1)} 	\ldots\!  {h}_{1 N_s}^{(1)} \\
			{h}_{1 1}^{(4)} \!\!	\ldots \!   {h}_{1 N_s}^{(4)} \\
			  \ddots    \\
			{h}_{\bar{N}_r 1}^{K^{(1)}}  \!\!	\ldots\! 	 {h}_{\bar{N}_r N_s} ^{K^{(1)}}
		\end{array}\!\!\right]\!\!,
		\tilde{\mathbf{H}}_y^{\mathrm{UC}}  \!\!\!=\!\! \left[\!\!
		\begin{array}{c}
			{h}_{1 1}^{(2)} \!\!	\ldots \!\!  {h}_{1 N_s}^{(2)} \\
			{h}_{1 1}^{(5)}  \!\!	\ldots \!\!   	 {h}_{1 N_s}^{(5)} \\
			  \ddots    \\
			\tilde{h}_{\bar{N}_r 1}^{K^{(2)}}   \!\! 	\ldots\!\!  	 {h}_{\bar{N}_r N_s} ^{K^{(2)}}
		\end{array}\!\!\right]\!\! , 
	\end{aligned}
\end{equation} 
\vspace{-0.1cm}
\vspace{-0.1cm}
\begin{equation}\notag
	\begin{aligned}
		\tilde{\mathbf{H}}_z^{\mathrm{UC}} &\!\!= \!\!\left[\!\!
		\begin{array}{c}
			{h}_{1 1}^{(3)}  \!\! 	\ldots   \!\!	 {h}_{1 N_s}^{(3)} \\
			{h}_{1 1}^{(6)} \!\! \ldots     \!\!	  {h}_{1 N_s}^{(6)} \\
		      \ddots   \\
			{h}_{\bar{N}_r 1}^{K^{(3)}}  \!\! 	\ldots   \!\!  	 {h}_{\bar{N}_r N_s} ^{K^{(3)}}
		\end{array}\right] \in \mathbb{C}^{K\bar{N}_r/3\times N_s}. 
	\end{aligned}
\end{equation}
\vspace{-0.1cm}
Since each user is assigned to one polarization, the cross-polarization and inter-user interference term are suppressed. The corresponding precoding matrices for the three polarizations are designed as follows $\forall$$i=1,2,\ldots,\bar{N}_r$:
\begin{equation}
	\begin{aligned}
		&\mathbf{P}_q= [p_{q,1}^{(1)},\ldots,p_{q,\bar{N}_r}^{(1)};\ldots;p_{q,1}^{(K)},\ldots,p_{q,\bar{N}_r}^{(K)}] , \\ 
		&\qquad \mathrm{where}\quad q\in\{x,y,z\}, \begin{cases}
			p_{q,i}^{(k)}=1, & \quad  k\in\mathcal{L}^q   \\
			p_{q,i}^{(k)}=0,  & \quad  \mathrm{otherwise}
		\end{cases}.
	\end{aligned}
\end{equation} 
It is noted that the proposed user-cluster-based precoding mitigates cross-polarization interference at the cost of system diversity. In fact, the system diversity is reduced since there is only one third of the patch antennas used for each user.  

\section{Numerical Results} \label{sec:simulation}
\begin{figure}  
	\begin{center}
		{\includegraphics[width=0.4 \textwidth]{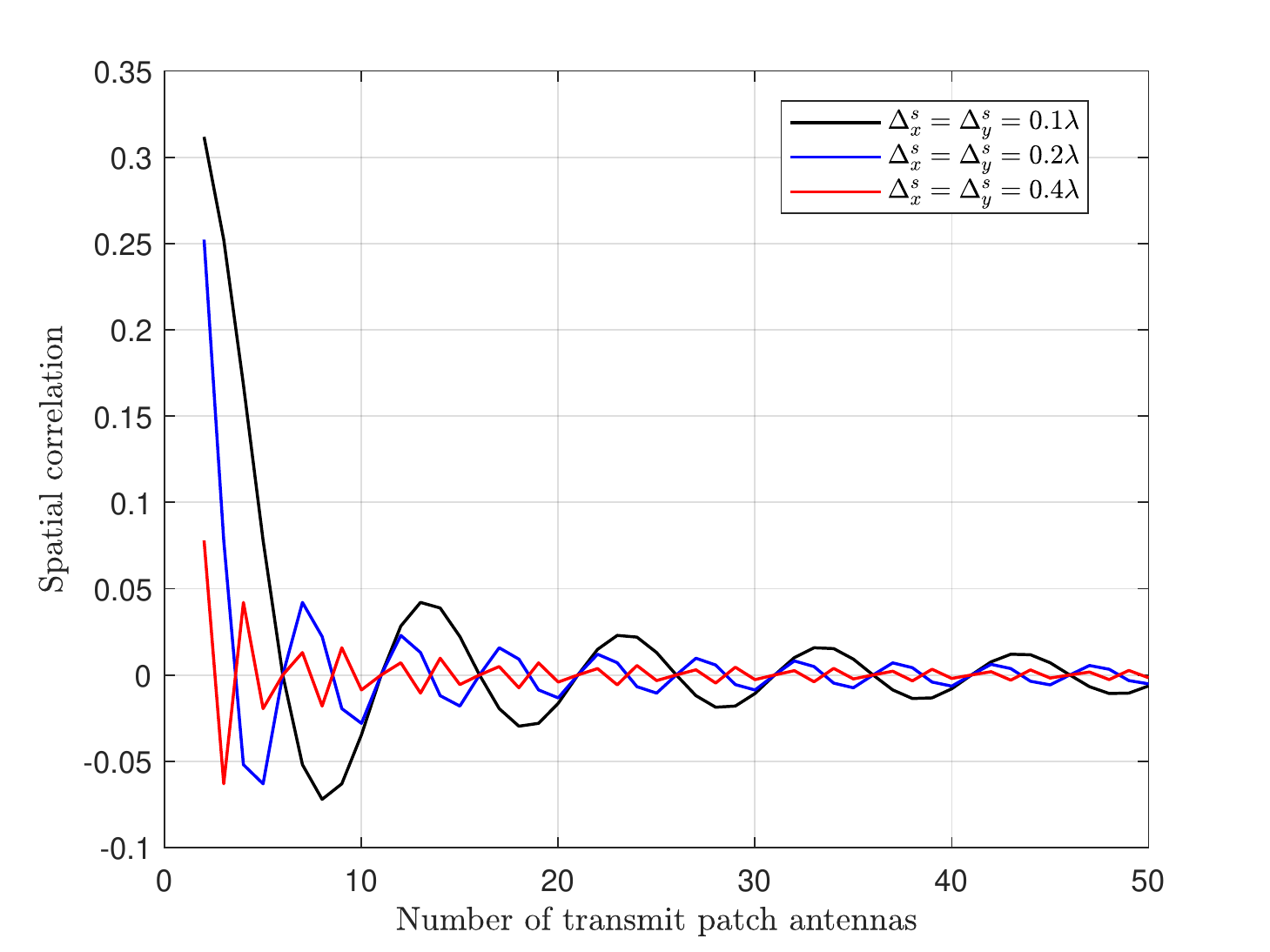}}  
		\caption{Simulated spatial correlation factors for different spacing versus the number of transmit antennas.}
		\label{fig:NorCorrelationUES} 
	\end{center}
\end{figure}

The transmit correlation factor $\mathbf{R}^s\left(\mathbf{r}_{n},\mathbf{r}_{n'}  \right)$ as a function of the number of transmit antennas is illustrated in Fig.~\ref{fig:NorCorrelationUES}. The wavelength $\lambda=1 \mathrm{m}$, the transmitter is equipped with $N_s=50$ patch antennas, and the spacing of adjacent transmit patch antennas is $\Delta^s_x=\Delta^s_y=0.1\lambda, 0.2\lambda$ and $0.4\lambda$.  It can be observed from the figure that, the larger number of transmit antennas reduces the normalized spatial correlation, since the distance between the two furthermost transmit patch antennas is increased. In addition, the larger spacing between patch antennas has the less correlation factor, which seems contradict to the theoretical analysis given by  \eqref{equ:CorrelationTaylor}. In fact, they all hold true since Fig.~\ref{fig:NorCorrelationUES} simulates the antennas with small size while  \eqref{equ:CorrelationTaylor} theoretically investigates infinitely small patch antennas, which is an ideal scenario. 

\begin{figure}  
	\begin{center}
		{\includegraphics[width=0.4 \textwidth]{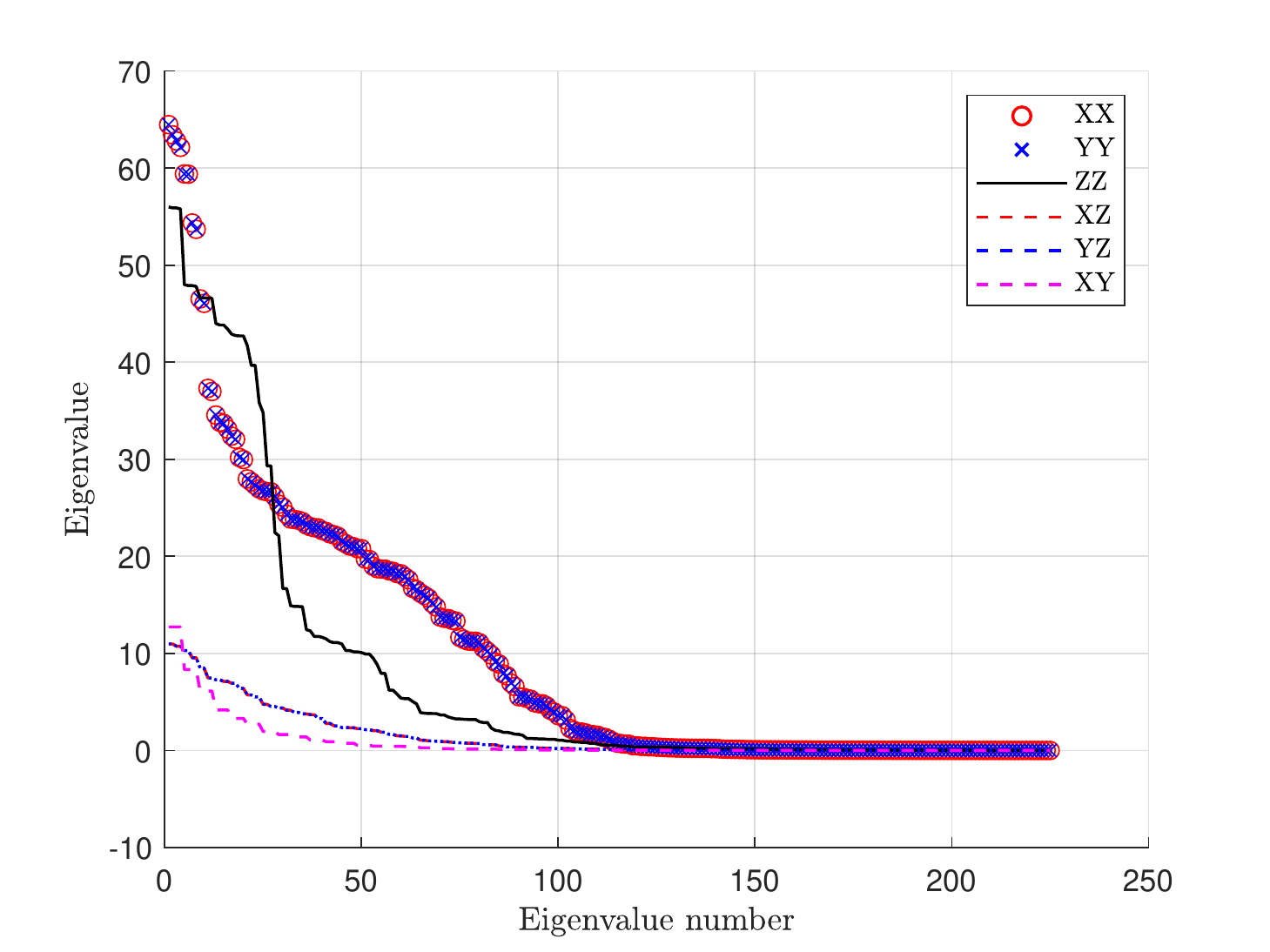}}  
		\caption{Eigenvalues of the co-polarized and cross-polarized HMIMOS wireless channels for a single user lying in the near-field regime.} 
		\label{fig:eigenvalueNF1lambda} 
	\end{center}
\end{figure}

The eigenvalue number of the co-polarized and cross-polarized channels $\mathbf{R}^{ij}=[\mathbf{H}^{ij}]^{\dagger} \mathbf{H}^{ij}, i,j\in\{x,y,z\}$ are depicted in Fig.~\ref{fig:eigenvalueNF1lambda}. The wavelength $\lambda=1 \mathrm{m}$, the transmitter and receiver are equipped with $225$ patch antennas, and the spacing between patch antennas is $0.4\lambda$. A single-user is located at the distance $z= \lambda$. It can be observed that the cross-polarization components are significant, thus, they need to be eliminated for efficient wireless communications. In addition, in the short distance, the non-zero eigenvalues of the $z$th co-polarized channel are as much as the other two co-polarized channels in Fig.~\ref{fig:eigenvalueNF1lambda}, implying similar contributes to the $x$ and $y$ co-polarized channels in the system.

\begin{figure}  
 	\begin{center}
 		{\includegraphics[width=0.4 \textwidth]{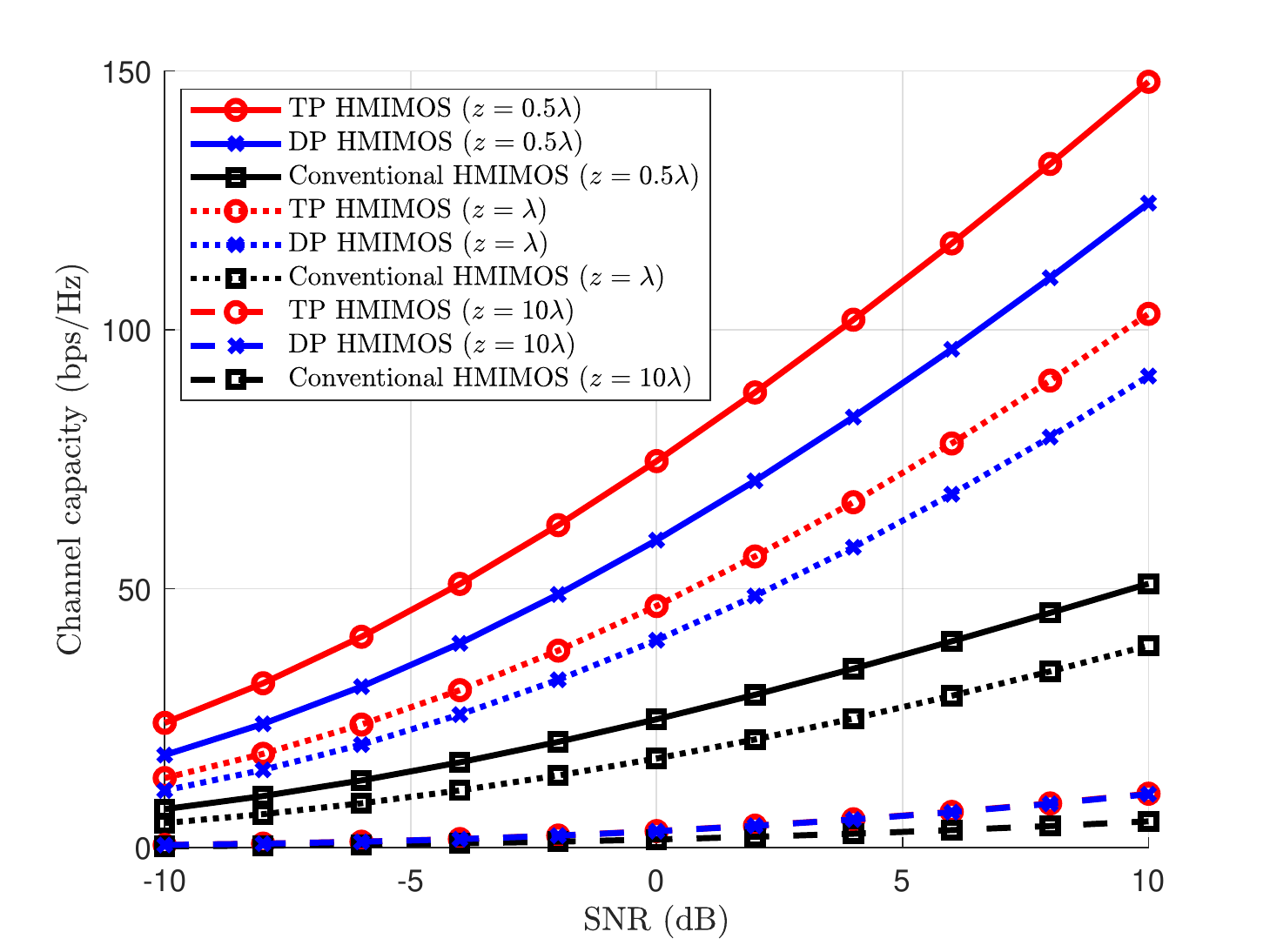}}  
 		\caption{The channel capacity of a TP HMIMOS, DP HMIMOS, and conventional HMIMOS systems. }
 		\label{fig:capacityTPDPSPIRS} 
 	\end{center}
 \end{figure}

The channel capacity comparison of TP HMIMOS, DP HMIMOS, and the conventional HMIMOS equipped with single polarized patch antennas is demonstrated in Fig.~\ref{fig:capacityTPDPSPIRS}. The wavelength $\lambda=1 \mathrm{m}$, the transmitter and receiver are equipped with $36$ and $9$ patch antennas, respectively, the spacing between patch antennas is $0.4\lambda$, and $K=3$ users are located at the distances $z=0.5 \lambda$, $\lambda$ and $10\lambda$, respectively. It can be observed that the TP HMIMOS has the largest capacity, since the full polarization is exploited. However, as the distance $z$ between the transmitter and receivers increases, the gap between TP HMIMOS and DP HMIMOS decreases. This showcases that the $z$th polarization component decays fast with the distance. It is thus apparent that the capacity of TP HMIMOS gradually coincides with the DP HMIMOS as the $z$ distance increases. 

\section{Conclusion}\label{sec:conclusion}
This paper presented a near-field channel model for TP MU-HMIMOS wireless communication systems, which was based on the dyadic Green's function. The proposed channel model was used to design a user-cluster-based precoding scheme for mitigating cross-polarization and inter-user interferences, which are indispensable components in polarized systems. 
Our simulation results showcased that TP HMIMOS systems have higher channel capacity than both DP and conventional HMIMOS in the near-field regime, however, this superiority gradually vanishes in the far-field regime as the $z$th polarized components disappear.   

\section{Acknowledgments}\label{sec:ack}
This work has been supported by the  China National  Key  R\&D  Program  under Grant No 2021YFA1000500, National Natural Science Foundation of China under Grant No 62101492, Zhejiang Provincial Natural Science Foundation of China under Grant No LR22F010002, National  Natural Science Fund for Excellent Young Scientists Fund Program (Overseas), Ng Teng Fong Charitable Foundation in the form of ZJU-SUTD IDEA Grant, Zhejiang University Education Foundation Qizhen Scholar Foundation, Fundamental Research Funds for the Central Universities under Grant No 2021FZZX001-21, the SNS JU TERRAMETA project under EU's Horizon Europe research and innovation programme under Grant Agreement No 101097101, and the Ministry of Education, Singapore, under its MOE Tier 2 (Award No MOE-T2EP50220-0019). Any opinions, findings and conclusions or recommendations expressed in this material are those of the authors and do not reflect the views of the Ministry of Education, Singapore.


\bibliographystyle{IEEEbib}
\bibliography{strings}

\end{document}